\documentclass{article}
\usepackage{spconf,amsmath,graphicx}
\usepackage{epstopdf}
\usepackage{array}
\usepackage{graphicx}
\usepackage{multirow}
\usepackage{color, soul}
\usepackage{cite}
\usepackage{url}
\usepackage{hyperref}
\usepackage{booktabs}
\usepackage{amsmath}

\usepackage{amssymb}

\title{StrengthNet: Deep Learning-based Emotion Strength Assessment for Emotional Speech Synthesis}

\name{Rui Liu$^{1,2}$, Berrak Sisman$^{2}$, Haizhou Li$^{1}$}
\address{$^1$ National University of Singapore $^2$ Singapore University of Technology and Design (SUTD) \\
\small{liurui\_imu@163.com, berraksisman@u.nus.edu, haizhou.li@nus.edu.sg}
}

\begin{document}
\maketitle
\begin{abstract}
Recently, emotional speech synthesis has achieved remarkable performance. The emotion strength of synthesized speech can be controlled flexibly using a strength descriptor, which is obtained by an emotion attribute ranking function. However, a trained ranking function on specific data has poor generalization, which limits its applicability for more realistic cases.
In this paper, we propose a deep learning based emotion strength assessment network for strength prediction that is referred to as StrengthNet.
Our model conforms to a multi-task learning framework with a structure that includes an acoustic encoder, a strength predictor and an auxiliary emotion predictor.
A data augmentation strategy was utilized to improve the model generalization.
Experiments show that the predicted emotion strength of the proposed StrengthNet are highly correlated with ground truth scores for seen and unseen speech.
Our codes are available at: \textcolor{blue}{\href{https://github.com/ttslr/StrengthNet}{https://github.com/ttslr/StrengthNet}}.


\end{abstract}
%


\section{Introduction}
\label{sec:intro}
Emotional speech synthesis (ESS), such as emotional text-to-speech, emotional voice conversion, etc., seeks to generate expressive speech with a desired emotion category.
Note that the control of emotion strength in an utterance is the key to emotion rendering.
Specifically, the emotion strength of synthesized speech can be controlled flexibly using a predefined strength descriptor. 

The simplest emotion strength control method is to weight the emotion representation vector by a scalar \cite{li2021controllable}, in which the weighting scalar holds poor interpretability.
To obtain a meaningful strength descriptor, some studies \cite{zhu2019controlling,lei2021fine} have tried to follow ``relative attributes'' \cite{parikh2011relative} and quantify the emotion strength based on the $<$neutral, emotional$>$ speech pairs.
Attribute ranking \cite{parikh2011relative} learns the difference between two samples that are significantly different in a particular attribute, that has been widely studied in computer vision \cite{siddiquie2011image,meng2018efficient}.
In speech processing, Zhu et al. \cite{zhu2019controlling} proposed to learn an emotion attribute ranking function $R(\cdot)$ from the $<$neutral, emotional$>$ paired speech features,
then weight the emotional feature with a learnable weighting vector and return a weighted sum $i$ indicating the emotion strength of one specific emotional speech.
Lei et al. \cite{lei2021fine} further extended the utterance level emotion attribute ranking function to phoneme level and obtain a fine-grained ranking function.
In this way, we can obtain a meaningful strength score for emotional speech, which correlates with human perception, of a specific dataset.

\begin{figure*}[!thp]
\centering
\centerline{ \quad \quad \includegraphics[width=\linewidth]{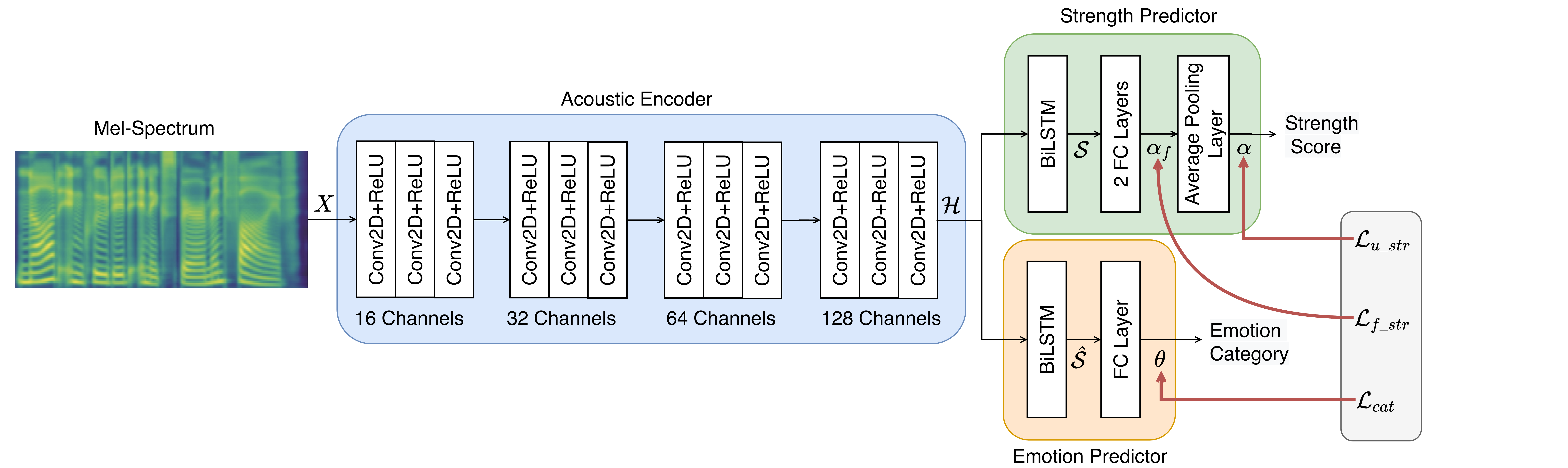}}
\vspace{-4mm}
\caption{The proposed StrengthNet that consists of Acoustic Encoder, Strength Predictor and Emotion Predictor.}
\label{fig:model}
\vspace{-3mm}
\end{figure*}

We note that a trained ranking function $R(\cdot)$ on specific data has poor generalization. In other words, a ranking function, $R(\cdot)$, trained on specific data can not calculate an exact strength score for unseen or our-of-domain speech. To match the new data, we need to retrain a new ranking function on the $<$neutral, emotional$>$ paired samples extracted from new data. We also note that the learning of ranking functions for new data is impossible without parallel samples. The above issues limit the applicability of speech emotion strength assessment for more realistic cases.
Recently, deep learning has shown its strong capacity to learn a mapping function in many different applications \cite{mosnet,qualitynet}.
Deep learning based networks can learn complex non-linear mapping relationships and can exhibit good generalization with the support of a large number of model parameters \cite{zhang2017Understanding}.

In this paper, we propose a novel deep learning based emotion strength assessment model, termed StrengthNet, which take a multi-task framework. StrengthNet consists of a convolutional neural network (CNN) based acoustic encoder, a bidirectional long short-term memory (BiLSTM) based strength predictor and an auxiliary BiLSTM based emotion predictor.
The acoustic encoder extracts the high-level features from the input mel-spectrum. The strength predictor aims to predict the strength score for input mel-spectrum. The emotion predictor was used to predict the emotion category, that serve as an auxiliary task. To improve the model generalization, we employ multiple emotional speech datasets and adopt a data augmentation strategy.

The main contributions of this paper are listed as follows:
1) We propose a novel deep learning based speech emotion strength assessment model, denoted as StrengthNet;
2) Experiments show that the predicted emotion strength of the proposed StrengthNet are highly correlated with ground truth scores for seen and unseen speech.
3) These results prove that the proposed StrengthNet can be used as a computational evaluator to assess the emotion strength for emotional speech synthesis.
To our best knowledge, this is the first deep learning-based emotion strength assessment model for emotional speech synthesis.

This paper is organized as follows:
In Section 2, we discuss the need for deep learning based emotion strength assessment to motivate our work.
In Section 3, we formulate the proposed StrengthNet. We report the experimental results in Section 4. Finally, Section 5 concludes the study.

\vspace{-2mm}
\section{The need for deep learning-based emotion strength assessment}
\vspace{-2mm}
At the present stage of emotional speech synthesis, an ESS model that is joint trained with other related models, such as speech emotion recognition \cite{cai2020emotion}, can achieve controllable ESS \cite{li2021controllable} or enhance the emotion expressiveness \cite{liu21p_interspeech}.
However, existing joint training models only consider the emotion category and ignore the emotion strength.
Traditional emotion strength prediction methods based on ranking functions only perform in the data pre-processing stage \cite{zhu2019controlling, lei2021fine}, and are impossible to integrate deeply in the training stage of ESS models.
Therefore, we would like to obtain a deep learning-based emotion prediction model that can be used as a front-end (aims for controllable) or back-end module (serve as perceptual loss \cite{9420276}, etc.) for emotional speech synthesis to enhance the performance in terms of emotion expressiveness.

We note that there are some attempts \cite{schnell11improving} trying to extract the emotion strength score from neural network based model to improve the ESS model. However, frame-level strength score extracted by unsupervised ways still lacks of interpretability.

\vspace{-3mm}
\section{StrengthNet}
\label{sec:StrengthNet}
\vspace{-3mm}
We propose a novel deep learning based speech emotion strength assessment model, denoted as StrengthNet. In this section, we first describe the overall StrengthNet architecture, then explain the model training details. Lastly, we describe the model inference.

\subsection{Model Overview}
\label{subsec:StrengthNet}

The overall model architecture is shown in the gray panel of Fig. \ref{fig:model}.
Our StrengthNet follows a multi-task framework, which consists of an acoustic encoder, a strength predictor and an emotion predictor.
We will discuss all components in order next.
\vspace{-3mm}
\subsubsection{Acoustic Encoder}
\vspace{-2mm}
The acoustic encoder takes the acoustic feature sequence, that is mel-spectrum in this work, as input to extract high-level feature representation.
The acoustic encoder consists of 12 convolution layers.
The strategy of stacking more convolutional layers to expand the receptive field of CNN has been widely used to model time series data and yield satisfactory performance \cite{mosnet}.
Given an input mel-spectrum sequence $X$, the CNN based acoustic encoder aims to extract a high-level feature $\mathcal{H}$,
The high-level feature $\mathcal{H}$ is then fed to two predictors to predict the emotion strength score and emotion category, respectively.
\vspace{-3mm}
\subsubsection{Strength Predictor}
\vspace{-2mm}
The strength predictor then reads the high-level feature representation to predict the emotion strength score.
Recent studies have confirmed the effectiveness of combining CNN and BiLSTM for classification \cite{mosnet}, and recognition \cite{9420276} tasks.

The strength predictor consists of a BiLSTM layer, two FC layers, following by an average pooling layer.
The BiLSTM layer takes the high-level feature $\mathcal{H}$ as input to output the hidden states $\mathcal{S}$ for each time step.
We then use two FC layers to regress frame-wise hidden states $\mathcal{S}$ into a frame level scalar $\alpha_{f}$ to indicate the strength score of each frame.
Finally, an average pooling layer is applied to the frame-level scores to obtain the utterance-level strength score $\alpha$.

To supervise the training of the strength predictor, we define a mean absolute error (MAE) loss $\mathcal{L}_{u\_str}$ behind the average pooling layer to force the predicted utterance-level strength score $\alpha$ close to the ground-truth value.

To improve the convergence of StrengthNet, we further define another MAE loss\cite{mosnet} $\mathcal{L}_{f\_str}$, denoted as frame-wise constraint, behind the last FC layer to minimize the difference between the predicted frame-level strength score $\alpha_{f}$ and the ground-truth strength.

\vspace{-2mm}
\subsubsection{Emotion Predictor}
\vspace{-2mm}
At the same time, the auxiliary emotion predictor is used to read the encoder output to predict the emotion category.

Similar to strength predictor, emotion predictor also consists of a BiLSTM layer and an addition softmax layer.
The BiLSTM summarizes the temporal information of high-level feature $\mathcal{H}$ into another latent states $\hat{\mathcal{S}}$.
Finally, the softmax layer converts the latent states $\hat{\mathcal{S}}$ to the output probability for all emotion categories.
Accordingly, we can obtain the predicted emotion category $\theta$.
We define a ``categorical\_crossentropy'' loss $\mathcal{L}_{cat}$ to restrain the emotion predictor.
We formulate the final objective function $\mathcal{L}_{final}$ for training the StrengthNet as: $ \mathcal{L}_{final} = \mathcal{L}_{f\_str} + \mathcal{L}_{u\_str} + \mathcal{L}_{cat}$.

\vspace{-2mm}
\subsection{Data Augmentation}
\vspace{-2mm}
To improve the model generalization, we employ a data augmentation strategy \cite{volpi2018generalizing,shorten2019survey}.
As shown in the blue panel of Fig. \ref{fig:model}, we employ multiple emotional speech datasets to train our StrengthNet.

Specifically, we assume that $\mathcal{D}_{1}$, $\mathcal{D}_{2}$, ..., $\mathcal{D}_{K}$ means K emotional speech dataset.
We  train the emotion attribute ranking function $R(\cdot)$ for all datasets, since they all provide <neutral, emotional> pairwise speech samples.

To achieve this, we first build two sets, that are $O$ and $S$, which contains ordered and similar paired samples respectively.
Specifically, for each dataset $\mathcal{D}_{k}$ ($k \in$ [1,K]), we pick up one sample from neutral speech and another sample from emotional speech to build the ordered set $O$.
We expect that the emotion strength of the emotional sample is higher than that of the neutral sample.
For the similar set $S$, we pick up two samples from neutral speech (or emotional speech). We assume that two samples from the same domain (neutral or emotional) have similar emotion strength.
Finally, we follow \cite{zhu2019controlling} and build a support vector machines (SVM) \cite{joachims2002optimizing} to learn the ranking function $R(\cdot)$ for emotion strength attribute.
After that, we extract the strength scores for all datasets as the ground-truth strength scalar which serve as the training objective of StrengthNet. Note that the strength score is normalized to (0,1) with 1 as the highest strength.
The mel-spectrum features for all datasets are treated as the input of StrengthNet.

The augmented data will represent a more comprehensive set, thus minimizing the distance between the training and validation set, as well as any future testing sets.

 \begin{figure}[!t]
\centering
\centerline{ \includegraphics[width=\linewidth]{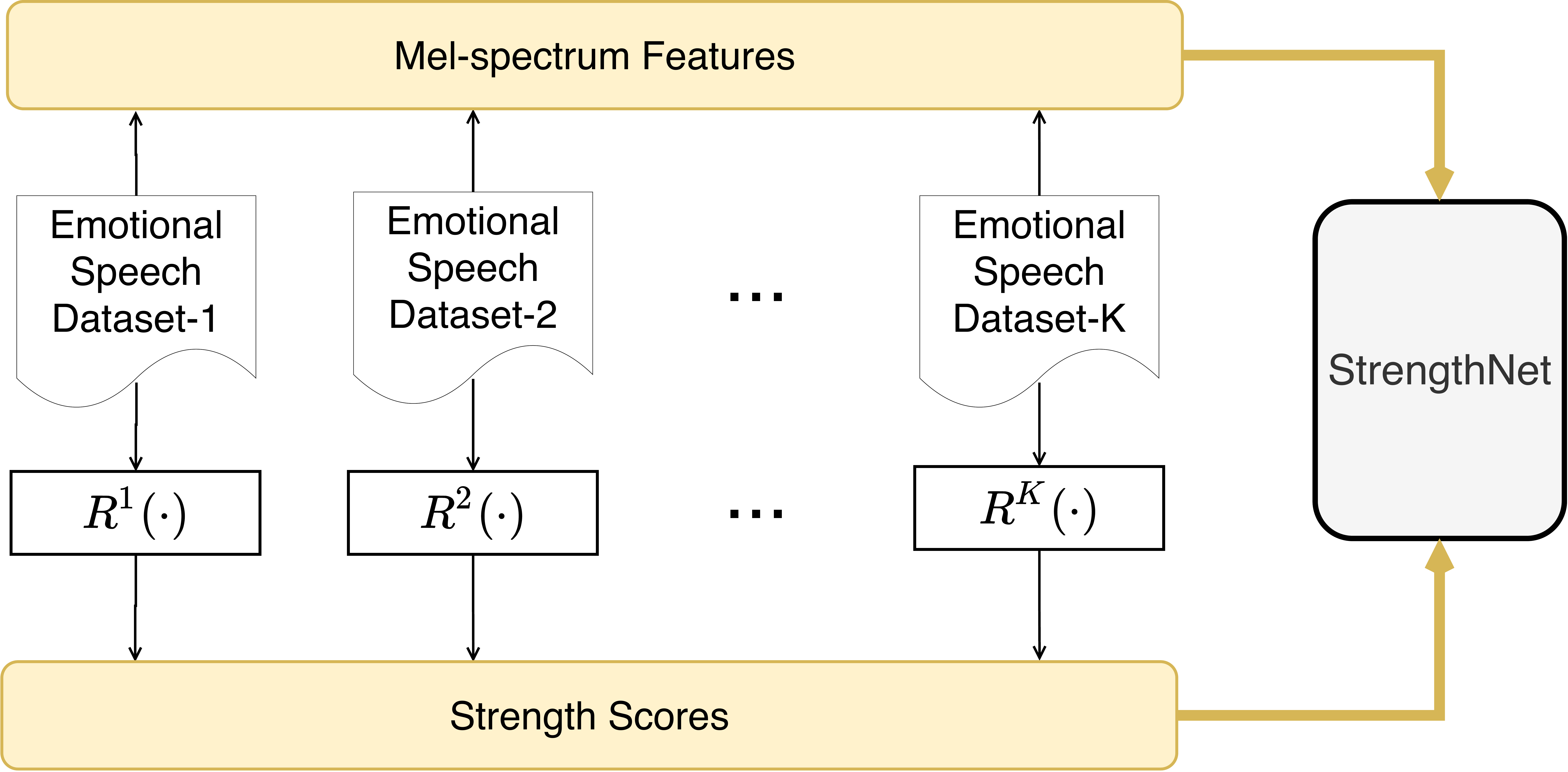}}
\vspace{-3mm}
\caption{The data augmentation strategy is utilized to improve the model generalization.}
\label{fig:dataaug}
\vspace{-4mm}
\end{figure}

 \vspace{-2mm}
\subsection{Run-time Inference}
\vspace{-2mm}
During inference, the StrengthNet takes a mel-spectrum extracted by any emotional speech as the input feature to predict its emotion strength score, as well as its emotion category.

Furthermore, our StrengthNet can be used directly to predict emotion strength for new emotional speech dataset without retraining, which we believe is remarkable.


\begin{figure*}[!t]
 \vspace{-3mm}
\centering
\centerline{
\includegraphics[width=0.33\linewidth]{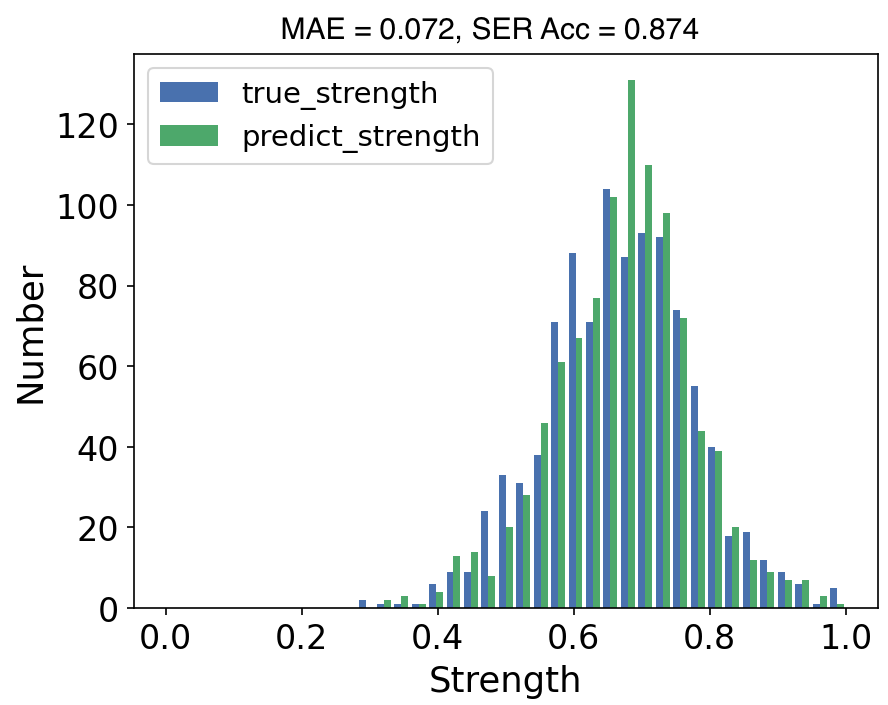}
\includegraphics[width=0.33\linewidth]{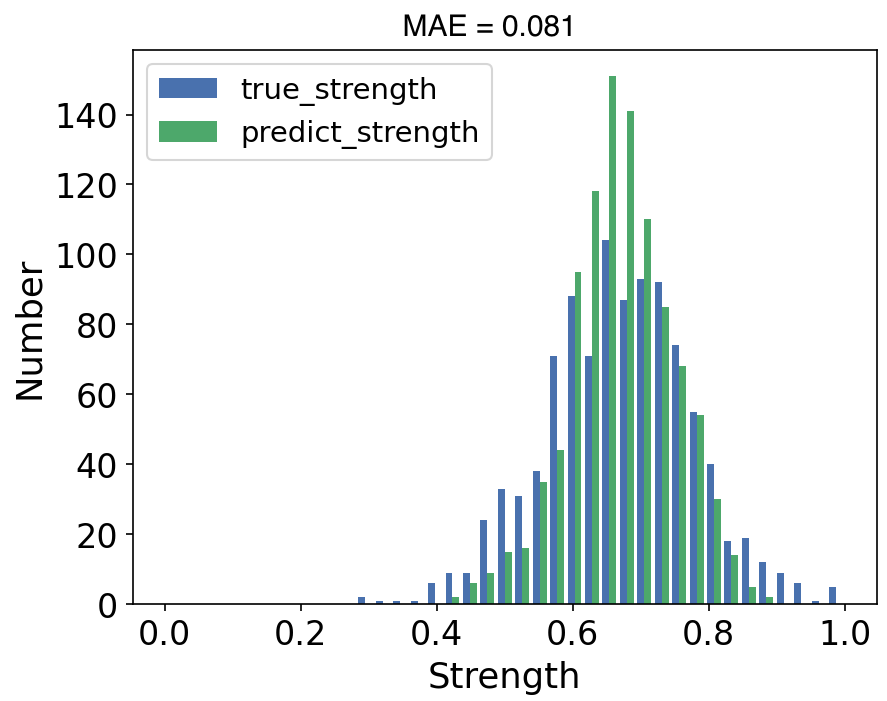}
\includegraphics[width=0.33\linewidth]{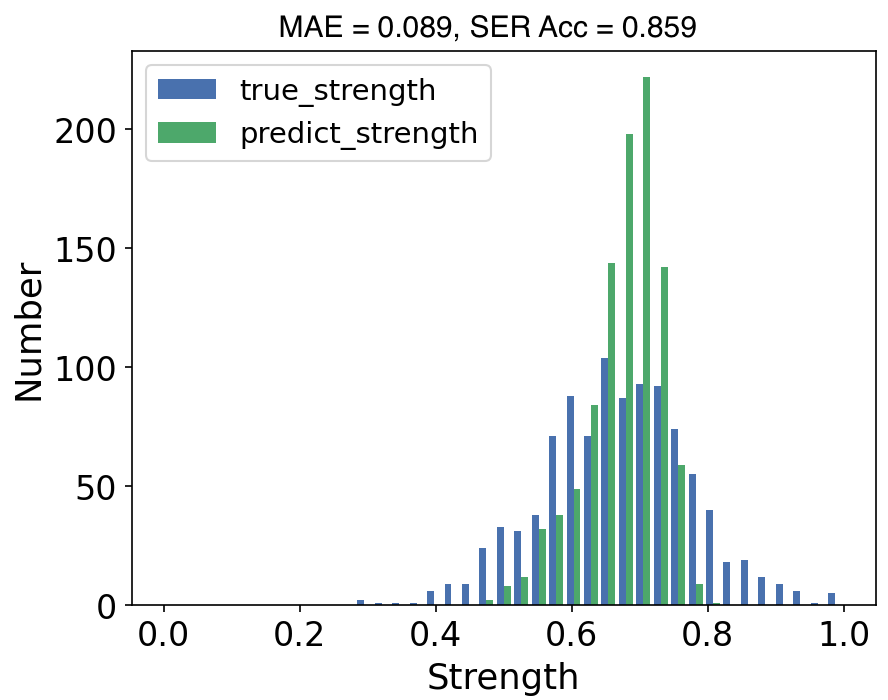}}

\centerline{\quad \quad \quad \quad (a) StrengthNet  \quad \quad \quad \quad \quad \quad \quad \quad   \quad  (b) StrengthNet w/o $\mathcal{L}_{cat}$ \quad \quad \quad \quad \quad  \quad  \quad  \quad  (c) StrengthNet w/o $\mathcal{L}_{f\_str}$ }
\vspace{-3mm}
\caption{Histogram of the utterance level strength predictions for (a) StrengthNet; (b) StrengthNet w/o $\mathcal{L}_{cat}$; (c) StrengthNet w/o $\mathcal{L}_{f\_str}$. The X-axis and Y-axis of subfigures represent the  strength scores and the utterance number, respectively.}
\label{fig:ablation_study}
\vspace{-3.5mm}
\end{figure*}

\section{Experiments}
\vspace{-3.5mm}
\subsection{Datasets}
\vspace{-2mm}
We use the ESD dataset \cite{zhou2021seen} to validate the performance of StrengthNet in terms of strength prediction. In addition, we use two additional SER datasets: Ryerson Audio-Visual Database of Emotional Speech and Song (RAVDESS) \cite{livingstone2018ryerson} and Surrey Audio-Visual Expressed Emotion (SAVEE) database \cite{jackson2014surrey} to achieve data augmentation and test the model generalization.






\vspace{-5mm}
\subsection{Experimental Setup}
\vspace{-2mm}

For each dataset, we take 5 emotion classes, that are happy, sad, angry, surprise, and neutral, to build the $<$neutral, emotional$>$ paired speech to train the ranking function.
During training of StrengthNet, we take 4 emotion classes, that are happy, sad, angry and surprise, into research.

We extract 80-channel mel-spectrum features with a frame size of 50ms and 12.5ms frame shift, that are further normalized to zero-mean and unit-variance, to serve as the model input.
The utterance level strength scores for all datasets are obtained using their ranking function and serve as the ground truth score, or training target, for strength predictor in our StrengthNet. We employ opensmile \cite{eyben2013recent} to extract 384-dimensional feature for each utterance for ranking function training.
To calculate the frame level MAE, the ground-truth strength score is used for all the frames in the speech utterance.
The emotion predictor of StrengthNet aims to predict four emotion categories, including happy, sad, angry and surprise.

The acoustic encoder consists of 4 Conv2D blocks with filters size [16, 32, 64, 128] respectively. Each block includes 3 Conv2D layers with strides shape \{[1,1], [1,1], [1,3]\} respectively. All layers share same kernel size [3 $\times$ 3] and ReLU activation function.
For BiLSTM layers in two predictors, each direction contains 128 cells.

All speech samples are resampled to 16 kHz.
We train the models using Adam optimizer with a learning rate=0.0001 and $\beta_1$ = 0.9, $\beta_2$ = 0.98. We set batch size to 64.
The dropout rate is set to 0.3.
For each dataset, we partition the speech data into training, validation, and test set at a ratio of 8:1:1.
We apply early stopping based on the MAE of the validation set with 30 epochs patience.

\subsection{Experimental Results}

\subsubsection{Architecture Comparison Results}
\vspace{-2mm}
First, we intend to validate the effectiveness of multi-task framework and frame constraint in our StrengthNet, in terms of strength prediction performance on ESD dataset. We develop StrengthNet and 2 variants for an ablation study, reported as follows:
1) StrengthNet (\textit{proposed}), which is our proposed model that consists of acoustic encoder, strength predictor and auxiliary emotion predictor;
2) StrengthNet w/o $\mathcal{L}_{cat}$, which is the proposed model without auxiliary emotion predictor; and
3) StrengthNet w/o $\mathcal{L}_{f\_str}$, which is the proposed model without frame strength constraint term $\mathcal{L}_{f\_str}$ within the strength predictor.

Fig. \ref{fig:ablation_study} shows the overall performance of these systems at the utterance level. We report the MAE value of emotion strength prediction and the accuracy of emotion category prediction on test data.
We observe in Fig. \ref{fig:ablation_study} that our proposed StrengthNet outperforms Strength w/o $\mathcal{L}_{cat}$ and Strength w/o $\mathcal{L}_{f\_str}$ and achieves the best performance, that is attributed to the multi-task and frame constraint strategy.  Specifically, we find that StrengthNet achieves the lowest MAE score of 0.072 and highest emotion recognition accuracy (denoted as \textit{SER Acc}) of 0.874.
To sum, the above results indicate that the combination of multi-task learning and frame constraint can effectively learn the emotion strength cues from the input mel-spectrum to perform emotion strength prediction well, along with category prediction.
\begin{table}
\centering
\caption{The MAE results for RAVDESS and SAVEE datasets in the comparison.}
\begin{tabular}{lcc}
\toprule
\multirow{2}{*}{\textbf{Method}}& \multicolumn{2}{c}{\textbf{MAE}}\\  
  &  RAVDESS & SAVEE  \\
 \bottomrule
$R_{{\rm RAVDESS}}(\cdot)$ & NA & 0.304 \\ \cline{1-3}
$R_{{\rm SAVEE}}(\cdot)$ & 0.283 & NA \\ \cline{1-3}
$R_{{\rm ESD}}(\cdot)$ & 0.266 & 0.272 \\ \cline{1-3}
$\rm StrengthNet_{{\rm ESD}}$ & 0.238 & 0.243 \\ \cline{1-3}
$\rm StrengthNet_{{\rm ESD+RAVDESS}}$ & NA & \textbf{0.173} \\ \cline{1-3}
$\rm StrengthNet_{{\rm ESD+SAVEE}}$ & \textbf{0.102} & NA \\
\bottomrule
\end{tabular}
\label{tab:dataaug}
 \vspace{-3mm}
\end{table}

\vspace{-3mm}
\subsubsection{Data Augmentation Results}
\vspace{-2mm}
We further verify the effectiveness of the proposed data augmentation for model generalization in terms of emotion strength prediction.

\begin{sloppypar}
We train StrengthNet on three dataset settings, that are ESD, ESD+RAVDESS, ESD+SAVEE.
$\rm  StrengthNet_{ESD}$, $\rm StrengthNet_{ESD+RAVDESS}$ and $\rm StrengthNet_{ESD+SAVEE}$ are used to represent three trained models respectively.
$R_{{\rm RAVDESS}}(\cdot)$, $R_{{\rm SAVEE}}(\cdot)$ and $R_{{\rm ESD}}(\cdot)$ refer to the three trained ranking functions for three datasets respectively.
We adopt all trained ranking functions and models to predict the emotion strength score on unseen test data from RAVDESS or SAVEE, and report the MAE results in Table \ref{tab:dataaug}.
\end{sloppypar}

As shown in Table \ref{tab:dataaug}, we observe that our StrengthNet performs lower MAE than ranking functions on both RAVDESS and SVAEE datasets. More importantly, in the case of $\rm StrengthNet_{ESD+RAVDESS}$ and $\rm  StrengthNet_{ESD+SAVEE}$, the MAE achieves the lowest values of 0.173 and 0.102 on SVAEE and RAVDESS respectively.
As can be seen from the results, our proposed strength method can reduce the overall MAE on unseen data with the data augmentation strategy, which performs better model generalization than attribute ranking function.

\vspace{-3mm}
\section{Conclusion}
 \vspace{-2mm}
This paper presents a deep learning-based speech emotion strength assessment model for the emotional speech synthesis task, referred to as StrengthNet.
Experimental results demonstrate that our StrengthNet can achieve accurate emotion strength prediction in a multi-task framework and frame constraint strategy. 
The model generalization also performs well with the help of the data enhancement strategy.
As per our knowledge, the proposed StrengthNet is the first end-to-end speech emotion strength assessment model.
In future work, we intend to integrate our StrengthNet as a front-end or back-end for emotional speech synthesis models to enhance the emotion expressiveness of output emotional speech.

 




\bibliographystyle{IEEEbib}
{\footnotesize
\bibliography{strings}

\begin{thebibliography}{10}

\bibitem{li2021controllable}
Tao Li, Shan Yang, Liumeng Xue, and Lei Xie,
\newblock ``Controllable emotion transfer for end-to-end speech synthesis,''
\newblock in {\em 2021 12th International Symposium on Chinese Spoken Language
  Processing (ISCSLP)}. IEEE, 2021, pp. 1--5.

\bibitem{zhu2019controlling}
Xiaolian Zhu, Shan Yang, Geng Yang, and Lei Xie,
\newblock ``Controlling emotion strength with relative attribute for end-to-end
  speech synthesis,''
\newblock in {\em 2019 IEEE Automatic Speech Recognition and Understanding
  Workshop (ASRU)}. IEEE, 2019, pp. 192--199.

\bibitem{lei2021fine}
Yi~Lei, Shan Yang, and Lei Xie,
\newblock ``Fine-grained emotion strength transfer, control and prediction for
  emotional speech synthesis,''
\newblock in {\em 2021 IEEE Spoken Language Technology Workshop (SLT)}. IEEE,
  2021, pp. 423--430.

\bibitem{parikh2011relative}
Devi Parikh and Kristen Grauman,
\newblock ``Relative attributes,''
\newblock in {\em 2011 International Conference on Computer Vision}. IEEE,
  2011, pp. 503--510.

\bibitem{siddiquie2011image}
Behjat Siddiquie, Rogerio~S Feris, and Larry~S Davis,
\newblock ``Image ranking and retrieval based on multi-attribute queries,''
\newblock in {\em CVPR 2011}. IEEE, 2011, pp. 801--808.

\bibitem{meng2018efficient}
Zihang Meng, Nagesh Adluru, Hyunwoo~J Kim, Glenn Fung, and Vikas Singh,
\newblock ``Efficient relative attribute learning using graph neural
  networks,''
\newblock in {\em Proceedings of the European conference on computer vision
  (ECCV)}, 2018, pp. 552--567.

\bibitem{mosnet}
Chen{-}Chou Lo, Szu{-}Wei Fu, Wen{-}Chin Huang, Xin Wang, Junichi Yamagishi,
  Yu~Tsao, and Hsin{-}Min Wang,
\newblock ``Mosnet: Deep learning-based objective assessment for voice
  conversion,''
\newblock in {\em Interspeech 2019, 20th Annual Conference of the International
  Speech Communication Association, Graz, Austria, 15-19 September 2019},
  Gernot Kubin and Zdravko Kacic, Eds. 2019, pp. 1541--1545, {ISCA}.

\bibitem{qualitynet}
Szu{-}Wei Fu, Yu~Tsao, Hsin{-}Te Hwang, and Hsin{-}Min Wang,
\newblock ``Quality-net: An end-to-end non-intrusive speech quality assessment
  model based on {BLSTM},''
\newblock in {\em Interspeech 2018, 19th Annual Conference of the International
  Speech Communication Association, Hyderabad, India, 2-6 September 2018},
  B.~Yegnanarayana, Ed. 2018, pp. 1873--1877, {ISCA}.

\bibitem{zhang2017Understanding}
Chiyuan Zhang, Samy Bengio, Moritz Hardt, Benjamin Recht, and Oriol Vinyals,
\newblock ``Understanding deep learning requires rethinking generalization,''
\newblock in {\em 5th International Conference on Learning Representations,
  {ICLR} 2017, Toulon, France, April 24-26, 2017, Conference Track
  Proceedings}. 2017, OpenReview.net.

\bibitem{cai2020emotion}
Xiong Cai, Dongyang Dai, Zhiyong Wu, Xiang Li, Jingbei Li, and Helen Meng,
\newblock ``Emotion controllable speech synthesis using emotion-unlabeled
  dataset with the assistance of cross-domain speech emotion recognition,''
\newblock in {\em ICASSP 2021-2021 IEEE International Conference on Acoustics,
  Speech and Signal Processing (ICASSP)}. IEEE, 2021, pp. 1--5.

\bibitem{liu21p_interspeech}
Rui Liu, Berrak Sisman, and Haizhou Li,
\newblock ``{Reinforcement Learning for Emotional Text-to-Speech Synthesis with
  Improved Emotion Discriminability},''
\newblock in {\em Proc. Interspeech 2021}, 2021, pp. 4648--4652.

\bibitem{9420276}
Rui Liu, Berrak Sisman, Guanglai Gao, and Haizhou Li,
\newblock ``Expressive tts training with frame and style reconstruction loss,''
\newblock {\em IEEE/ACM Transactions on Audio, Speech, and Language
  Processing}, vol. 29, pp. 1806--1818, 2021.

\bibitem{schnell11improving}
Bastian Schnell and Philip~N Garner,
\newblock ``Improving emotional tts with an emotion intensity input from
  unsupervised extraction,''
\newblock in {\em Proc. 11th ISCA Speech Synthesis Workshop (SSW 11)}, pp.
  60--65.

\bibitem{volpi2018generalizing}
Riccardo Volpi, Hongseok Namkoong, Ozan Sener, John Duchi, Vittorio Murino, and
  Silvio Savarese,
\newblock ``Generalizing to unseen domains via adversarial data augmentation,''
\newblock in {\em Proceedings of the 32nd International Conference on Neural
  Information Processing Systems}, 2018, pp. 5339--5349.

\bibitem{shorten2019survey}
Connor Shorten and Taghi~M Khoshgoftaar,
\newblock ``A survey on image data augmentation for deep learning,''
\newblock {\em Journal of Big Data}, vol. 6, no. 1, pp. 1--48, 2019.

\bibitem{joachims2002optimizing}
Thorsten Joachims,
\newblock ``Optimizing search engines using clickthrough data,''
\newblock in {\em Proceedings of the eighth ACM SIGKDD international conference
  on Knowledge discovery and data mining}, 2002, pp. 133--142.

\bibitem{zhou2021seen}
Kun Zhou, Berrak Sisman, Rui Liu, and Haizhou Li,
\newblock ``Seen and unseen emotional style transfer for voice conversion with
  a new emotional speech dataset,''
\newblock in {\em ICASSP 2021-2021 IEEE International Conference on Acoustics,
  Speech and Signal Processing (ICASSP)}. IEEE, 2021, pp. 920--924.

\bibitem{livingstone2018ryerson}
Steven~R Livingstone and Frank~A Russo,
\newblock ``The ryerson audio-visual database of emotional speech and song
  (ravdess): A dynamic, multimodal set of facial and vocal expressions in north
  american english,''
\newblock {\em PloS one}, vol. 13, no. 5, pp. e0196391, 2018.

\bibitem{jackson2014surrey}
Philip Jackson and SJUoSG Haq,
\newblock ``Surrey audio-visual expressed emotion (savee) database,''
\newblock {\em University of Surrey: Guildford, UK}, 2014.

\bibitem{eyben2013recent}
Florian Eyben, Felix Weninger, Florian Gross, and Bj{\"o}rn Schuller,
\newblock ``Recent developments in opensmile, the munich open-source multimedia
  feature extractor,''
\newblock in {\em Proceedings of the 21st ACM international conference on
  Multimedia}, 2013, pp. 835--838.

\end{thebibliography}
}
\end{document}